\pdfoutput=1
\documentclass[11pt]{article}

\usepackage[preprint]{acl}

\usepackage{times}
\usepackage{latexsym}

\usepackage[T1]{fontenc}
\usepackage[utf8]{inputenc}

\usepackage{microtype}

\usepackage{inconsolata}

\usepackage{graphicx}

\usepackage{algorithm}
\usepackage{algpseudocode}

\title{LLM-as-a-Judge for Reference-less Automatic Code Validation and Refinement for Natural Language to Bash in IT Automation}

\author{Ngoc Phuoc An Vo \\
  IBM Research \\
  Yorktown Heights, US \\
  \texttt{ngoc.phuoc.an.vo@ibm.com} \\\And
  Brent Paulovicks \\
  IBM Research \\
  Yorktown Heights, US \\
  \texttt{ovicks@us.ibm.com} \\\And
  Vadim Sheinin \\
  IBM Research \\
  Yorktown Heights, US \\
  \texttt{vadims@us.ibm.com} \\}

\begin{document}
\maketitle
\begin{abstract}
In an effort to automatically evaluate and select the best model and improve code quality for automatic incident remediation in IT Automation, it is crucial to verify if the generated code for remediation action is syntactically and semantically correct and whether it can be executed correctly as intended. There are three approaches: 1) \textbf{conventional methods} use surface form similarity metrics (token match, exact match, etc.) which have numerous limitations, 2) \textbf{execution-based evaluation} focuses more on code functionality based on pass/fail judgments for given test-cases, and 3) \textbf{LLM-as-a-Judge} employs LLMs for automated evaluation to judge if it is a correct answer for a given problem based on pre-defined metrics. In this work, we focused on enhancing LLM-as-a-Judge using bidirectional functionality matching and logic representation for reference-less automatic validation and refinement for Bash code generation to select the best model for automatic incident remediation in IT Automation. We used execution-based evaluation as ground-truth to evaluate our LLM-as-a-Judge metrics. Results show high accuracy and agreement with execution-based evaluation (and up to 8\% over baseline). Finally, we built Reflection code agents to utilize judgments and feedback from our evaluation metrics which achieved significant improvement (up to 24\% increase in accuracy) for automatic code refinement.

\end{abstract}

\section{Introduction}

An automatic incident remediation pipeline for IT Automation usually consists of following steps: 1) \textbf{diagnosis} where observability data is used to detect events and incidents, 2) \textbf{action recommendation} where remediation actions are generated for specific incidents, and 3) \textbf{action automation} where code scripts (i.e. Bash, PowerShell, Ansible, etc) are generated for remediation actions and applied to resolve the given incidents. In the scope of this paper, we focus on \textbf{action automation} step. Natural language to Bash (NL2Bash) is widely used to generate Bash scripts from natural language prompts for automating various tasks in IT Automation such as performance monitoring, compilation, system administration, system diagnostics, computing, etc. The important step is verifying the generated code as syntactically and semantically correct and whether it can be executed correctly as intended. In light of automating remediation actions for incidents discovered from various platforms in Application Performance Management (APM) tools (e.g. Dynatrace, AppDynamics), the main idea is to find the best model for generating code scripts. There are different ways to assess the quality of the code such as manual verification by a human or automatic verification using various evaluation metrics. Conventional methods for automatically evaluating code quality heavily rely on surface form similarity metrics such as BLEU \cite{papineni2002bleu}, ROUGE \cite{lin2004rouge}, and exact/partial match. This requires references for comparison and may overlook code syntax features, ignore semantics features and execution effects, or even over-penalize alternative solutions \cite{yang2023intercode}. In contrast, execution-based evaluation focuses more on code functionality and does not restrict the code generation to any fixed solution. Nevertheless, execution-based evaluation, especially for Bash is rather complex and costly to design and implement since the verification should take many factors into account (e.g. changes of test environment, environment status at different times, etc) and it only can be used offline. Lately, given recent advances of Large Language Models (LLMs), LLM-as-a-Judge paradigm employing LLMs for automated evaluation has emerged. Given strong coding abilities and reasoning skills, LLMs have the potential to be cost-effective and scalable surrogates for human evaluators. However, this research direction is still in its early stages with many needed breakthroughs. In this paper, we introduce \texttt{bidirectional functionality matching} and \texttt{logic representation} metrics to enhance LLM-as-a-Judge for reference-less automatic quality assessment and refinement for Bash. \texttt{Bidirectional functionality matching} validates if the functionality description of a code snippet satisfies and covers all required functionalities of the given problem and vice-versa. In contrast, regardless of complex programming syntax, a logic representation is generated which extracts the main logic of the given code snippet and then verified as a solution for the given problem. Our metrics show better agreement (in Accuracy) with execution-based evaluation than other metrics for LLM-as-a-Judge. Unlike others, ours not only produce a rating score or binary judgment to conclude if the code snippet is a correct answer for the given problem, but also provides evaluation details and feedback to possibly repair the incorrect code. Next we built Reflection code agents in which we incorporate judgments and feedback from our metrics to attempt to refine the original code. Results show significant improvements for code refinement across Bash datasets and models.

\section{Related Works}
With the continuous progress of Large Language Models (LLMs), their role in evaluating code generation has become an area of growing interest. Traditional evaluation approaches, such as BLEU, ROUGE, and execution-based methods, often fall short in comprehensively assessing both the correctness and overall quality of generated code. Recently, the "LLM-as-a-Judge" paradigm has gained traction, where LLMs are leveraged as evaluators to judge the quality of generated code across multiple dimensions, including correctness, readability, efficiency, and alignment with human judgment. This review examines recent advancements in utilizing LLMs for code evaluation, emphasizing key methodologies, benefits, and challenges.

\paragraph{Conventional Evaluation Methods} Conventional approaches to evaluating code generation primarily depend on: 1) surface similarity metrics such as BLEU, ROUGE, METEOR \cite{banerjee2005meteor}, CodeBLEU \cite{ren2020codebleu}, etc. quantify the overlap between generated and reference code at the n-gram level but do not effectively capture logical accuracy or intended functionality, 2) execution-based metrics runs generated code against test cases which helps determine functional correctness but does not evaluate aspects like efficiency, code structure, or adaptability to different contexts, and 3) human evaluation using expert assessment remains the most reliable approach but its scalability is limited due to high time and resource demands. Due to these constraints, researchers have increasingly explored the potential of LLMs as automated judges, offering a more scalable and comprehensive solution for evaluating generated code. All these works require either references for comparison or expensive labeled data for training or costly execution-based platform that only can run offline.

\paragraph{LLM-as-a-Judge} Multiple studies have investigated how LLMs can serve as evaluators for code generation, surpassing conventional assessment techniques. These advancements seek to improve evaluation consistency while reducing reliance on human reviewers. GPTScore \cite{fu2023gptscore} utilizes LLMs to assign scores to generated code by analyzing semantic similarity and logical accuracy. G-EVAL \cite{liu2023g} leverages generative models for a comprehensive evaluation that includes readability, maintainability, and functional soundness. CodeBERTScore \cite{zhou2023codebertscore} fine-tunes LLMs with expert-annotated evaluation datasets which enhances their ability to differentiate between high and low quality code for specific programming languages (Python, Javascript, C, C++, Java); however, it relies on high-quality references that can be difficult and expensive to obtain. ICE-Score \cite{zhuo2024ice} leveraged G-EVAL and focused on assessing usefulness and functional correctness of generated code and proved to work effectively on four programming languages (Java, Python, C, C++, and JavaScript). Another work, CodeSift \cite{10643933}, attempted to compute similarities and differences between the given task description and code functionality for semantic correctness assessment.

Unlike these previous works, our metrics are: 1) reference-less and able to run online, 2) validating both functionality and logic correctness of the code snippet, 3) not only using the given problem as-is but also analyzing it to obtain all required functionalities to compare with the code snippet, and 4) comparing between the code snippet and the given problem in a bidirectional fashion.

\section{Bidirectional Functionality Matching vs. Logic Representation Metrics}
\label{sec:llm-judge}
To validate if the generated code snippet is a correct answer to the given problem, we introduced the following LLM-as-a-Judge reference-less metrics.

\subsection{Bidirectional Functionality Matching} A comprehensive functionality description refers to set of features or capabilities of a software application to perform specific tasks or functions. It encompasses the features and capabilities that enable users to achieve their desired objectives efficiently and effectively. Unlike other metrics computing only similarities or differences between the given problem and code functionality, we extract both 1) required functionalities of the given problem, and 2) comprehensive functionality description of the generated code snippet; then we compare if (2) totally satisfies and aligns with (1) and vice-versa. Consequently, we decide if the code snippet is a correct answer for the given problem (see Algorithm \ref{code_func}). Figure \ref{fig-func-des} shows an example of how a functionality description is used for validating the code snippet.

\begin{algorithm}
\caption{Validation via Bidirectional Functionality Matching}
\label{code_func}
\begin{algorithmic}[1]

\State \textbf{Step 1:} Extract required functionalities of given problem
\State $required\_func \gets extract\_prob\_func(problem)$

\State \textbf{Step 2:} Generate a comprehensive functionality description of the code snippet
\State $code\_func\_desc \gets GenerateFuncDesc(code\_snippet)$

\State \textbf{Step 3:} Compare functionality description with the problem's required functionalities
\State $is\_matched \gets Compare(code\_func\_desc, required\_func)$

\If {$is\_matched = true$}
    \State \textbf{correct functionality}
\Else
    \State \textbf{incorrect functionality}
\EndIf

\end{algorithmic}
\end{algorithm}

\subsection{Validation via Logic Representation}
To escape the burden of complex programming syntax, a logic representation in Pseudocode only focuses on the main logic of the given code snippet. We execute the following procedure: 1) extract required functionalities of the given problem, 2) translate the code snippet to a logic representation, 3) validate if the logic representation totally satisfies and covers the required functionalities. We use this to verify if the code snippet is a correct solution for the given problem (see Algorithm \ref{logic_rep}). Figure \ref{fig-logic-rep} shows an example of how a logic representation in Pseudocode is used for validating the code snippet.

\begin{algorithm}
\caption{Logic representation Validation}
\label{logic_rep}
\begin{algorithmic}[1]

% \State \textbf{Step 1:} Generate a logic representation to covers all required functionalities of the given problem
% \State $LR_1 \gets AnswerInLogicRep(problem)$
\State \textbf{Step 1:} Extract required functionalities of given problem
\State $required\_func \gets extract\_prob\_func(problem)$

\State \textbf{Step 2:} Translate the code snippet into logic representation
\State $LR \gets Translate2LogicRep(code\_snippet)$

\State \textbf{Step 3:} Validate if the logic rep satisfies and covers required functionalities of the problem
\State $is\_covered \gets cover(LR, required\_func)$

\If {$is\_covered = true$}
    \State \textbf{correct logic}
\Else
    \State \textbf{incorrect logic}
\EndIf

\end{algorithmic}
\end{algorithm}

\section{Experiments}
\label{sec:experiments}
\paragraph{Dataset} 
We used NL2Bash-EAbench\footnote{https://github.com/IBM/nl2bash-eabench/tree/main} dataset \cite{vo2024executionbasedevaluationnaturallanguage} consisting of three test suites and validators for execution-based evaluation of Bash for automatic incident remediation in IT Automation: 1) Bash-1 has 50 generic test cases using single-line Bash commands, 2) Bash-2 has 50 system-oriented test cases using single-line Bash commands (for performance monitoring, system diagnostics, system administration, etc), and 3) Bash-3 has 50 advanced test cases focusing on Bash-language constructs rather than the Linux commands with which it is normally conflated.

\paragraph{Processing for Automatic Code Extraction} We implemented a heuristic search algorithm to automatically extract the Bash code snippet from the output of LLMs. Most LLMs follow instructions quite well to only return, or to mark, the desired code snippet for the first round of code generation. However, after the reflection, quite often LLMs return the desired code snippets mixed with various reasoning, explanation, example texts (including hallucinations) without clear markings; thus we have to manually examine and extract the code.

To compare our metrics with others in the literature, we set up experiments for two tasks:

\subsection{Task 1: Automatic Code Validation}
To evaluate our metrics for automatic code assessment for Bash, we 
used the following setting:
\begin{enumerate}
    \itemsep0em
    \item Use \texttt{Granite-34b-code-instr} model\footnote{https://huggingface.co/ibm-granite/granite-34b-code-instruct-8k/tree/main} to produce predictions for questions in three test suites.
    \item Run execution-based evaluation on the predictions (1) and use output as ground-truth.
    \item Run our metrics and baseline to evaluate predictions (1).
    \item Compare evaluation outputs of our metrics and baseline against the ground-truth (2).
\end{enumerate}

And other configurations as below:
\begin{itemize}
    \itemsep0em
    \item \textbf{Pre-processing}: generated code snippets are always validated for syntactic correctness by the ShellCheck\footnote{https://www.shellcheck.net/} tool.
    \item \textbf{Baseline}: we used \texttt{ICE-Score} \cite{zhuo2024ice}, one of the latest and most advanced LLM-as-a-Judge metrics.
    \item \textbf{Goal}: to evaluate the accuracy and agreement of validation decisions between our metrics, the baseline, and the ground-truth.
    \item \textbf{Evaluation metrics}: since we used execution-based evaluation result as ground-truth which has binary decisions of pass vs. fail, we considered this as a classification problem, thus we used the following evaluation metrics: Accuracy, Precision, Recall, and F1-score.
\end{itemize}

\subsection{Task 2: Automatic Code Refinement}
Unlike other LLM-as-a-Judge metrics, ours not only produce a rating score or binary judgment to decide if the code snippet is a correct answer for the given problem, but also provide a detailed evaluation and feedback to possibly repair the incorrect code. To measure how much our metrics can contribute towards improving quality of the original generated code, we built two types of Reflection code agent to examine it. To build agents for various tasks (code generation, code assessment, and code refinement), we used CrewAI encapsulating ReAct framework\footnote{https://www.crewai.com/} and several top, open-source and closed-source LLMs: \texttt{Llama-3.3-70B}\footnote{https://github.com/meta-llama/llama-models/blob/main/models/llama3\_3/MODEL\_CARD.md}, \texttt{Mistral-large}\footnote{https://mistral.ai/news/mistral-large}, and \texttt{GPT-4o} \cite{openai2024gpt4technicalreport}.

\begin{figure}[htbp]
\centerline{\includegraphics[scale=0.55]{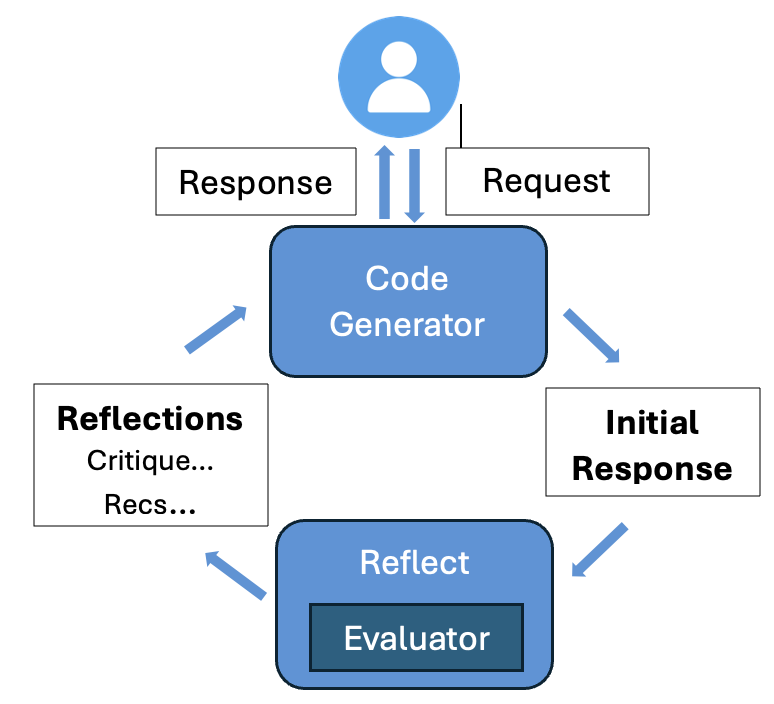}}
\caption{A Reflection Agent with Dedicated Evaluator}
\label{fig-agent}
\end{figure}

\paragraph{Self-Reflection agent}: A self-reflection agent can evaluate its actions and responses, then find areas for improvements, and consequently improve its strategies to obtain better results. After the \texttt{Code generator} has generated the initial code snippet based on the given problem (from user's request), it will be self-reviewed and self-evaluated by the critic (\texttt{Reflect}). If the initial response is not a correct answer to the given problem, the \texttt{Code generator} will attempt to refine the initial code snippet based on the self-evaluation and feedback (Figure \ref{fig-agent} but without the \texttt{Evaluator} module).
    
\paragraph{Reflection agent with dedicated Evaluator}: This is similar to the \texttt{Self-Reflection agent} but it includes a dedicated \texttt{Evaluator}. The initial code snippet (generated by \texttt{Code generator}) is evaluated by the \texttt{Evaluator} using the two LLM-as-a-Judge metrics (Section \ref{sec:llm-judge}), then the judgment and feedback are sent back to the \texttt{Code generator} for potential improvement (Figure \ref{fig-agent}).

\begin{figure*}[htbp]
\centerline{\includegraphics[scale=0.23]{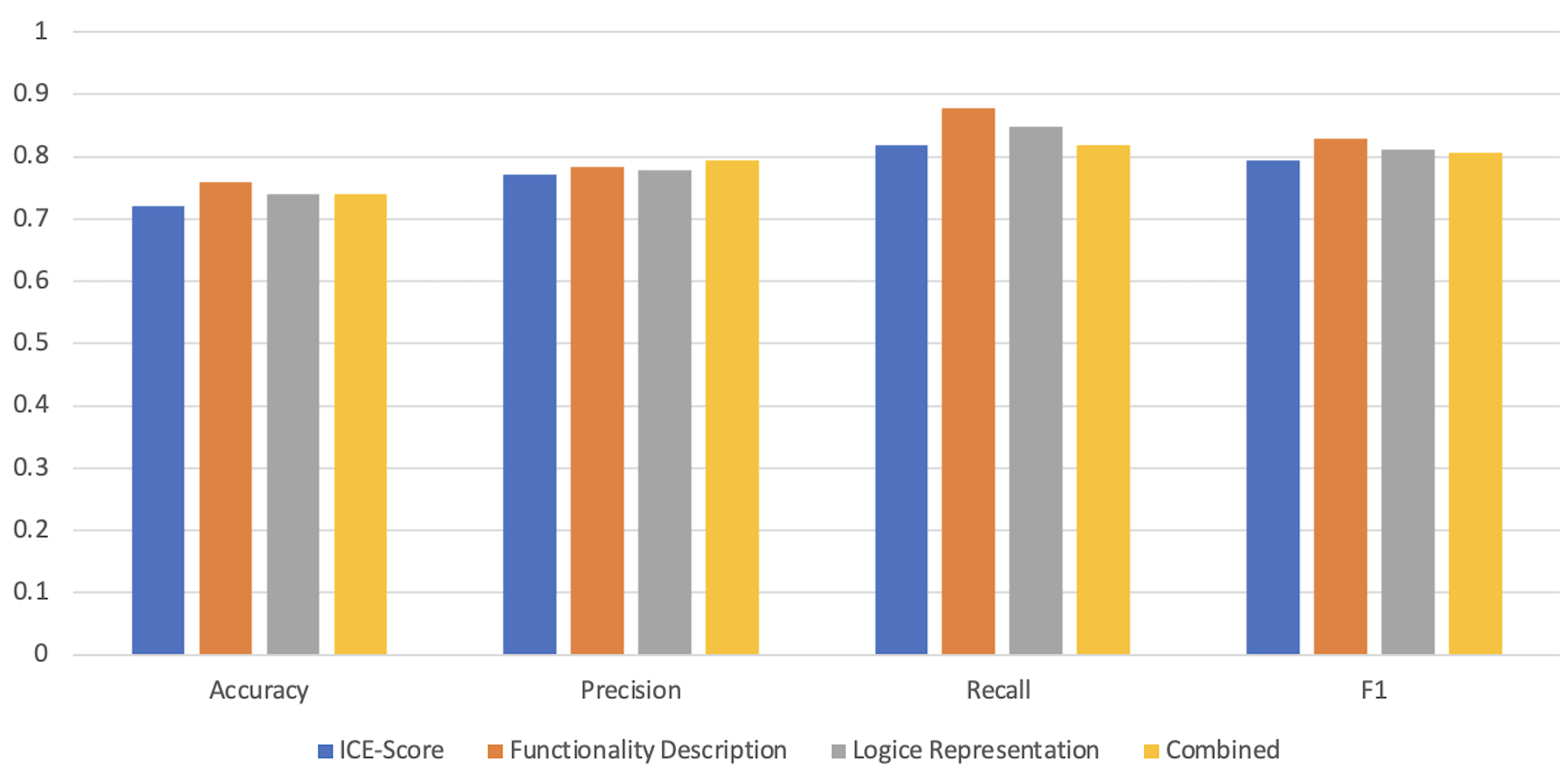}}
\caption{Comparison among LLM-as-a-Judge Metrics for Bash-1}
\label{fig-bash1}
\end{figure*}

\begin{figure*}[htbp]
\centerline{\includegraphics[scale=0.23]{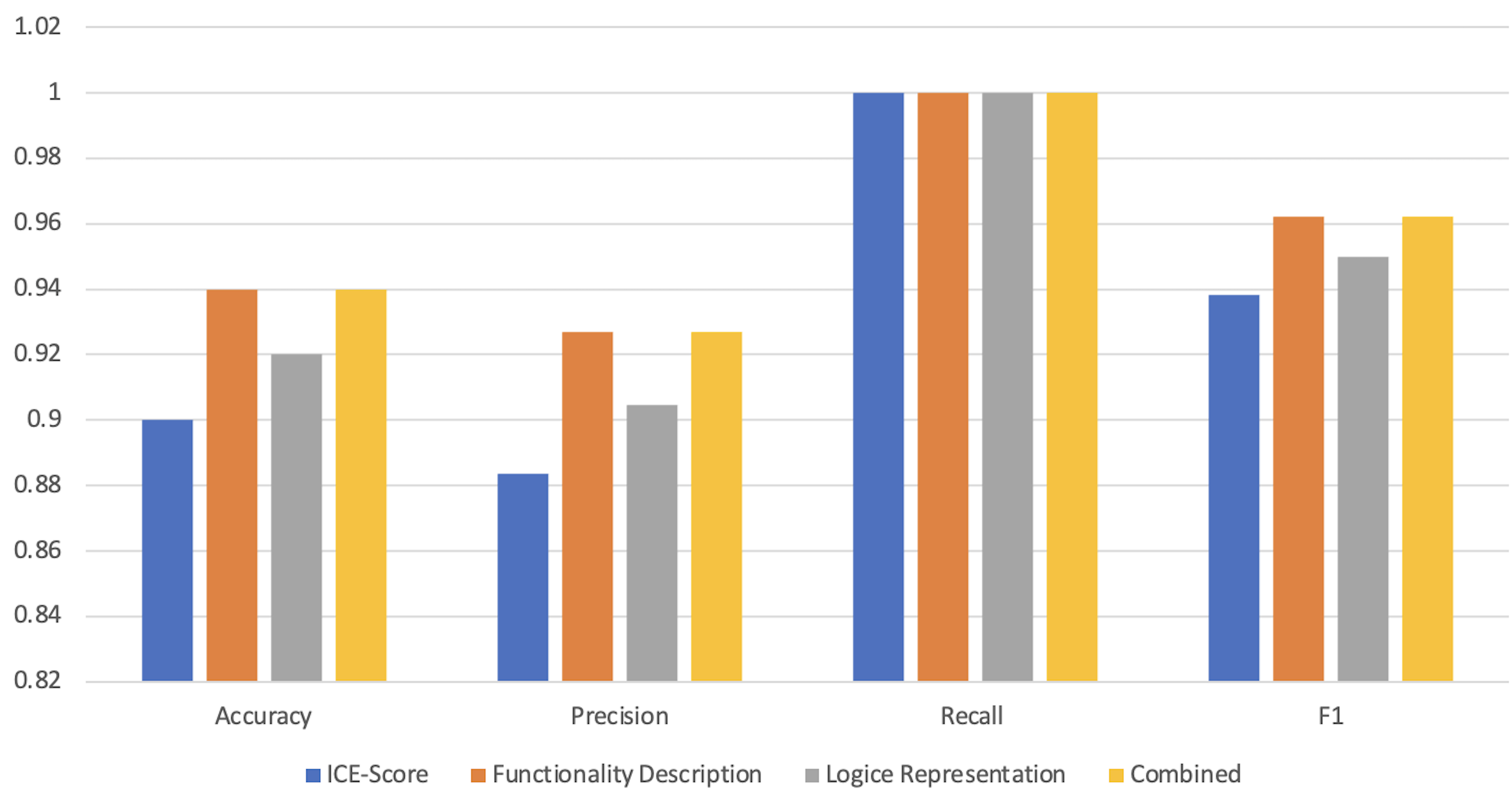}}
\caption{Comparison among LLM-as-a-Judge Metrics for Bash-2}
\label{fig-bash2}
\end{figure*}

\begin{figure*}[htbp]
\centerline{\includegraphics[scale=0.23]{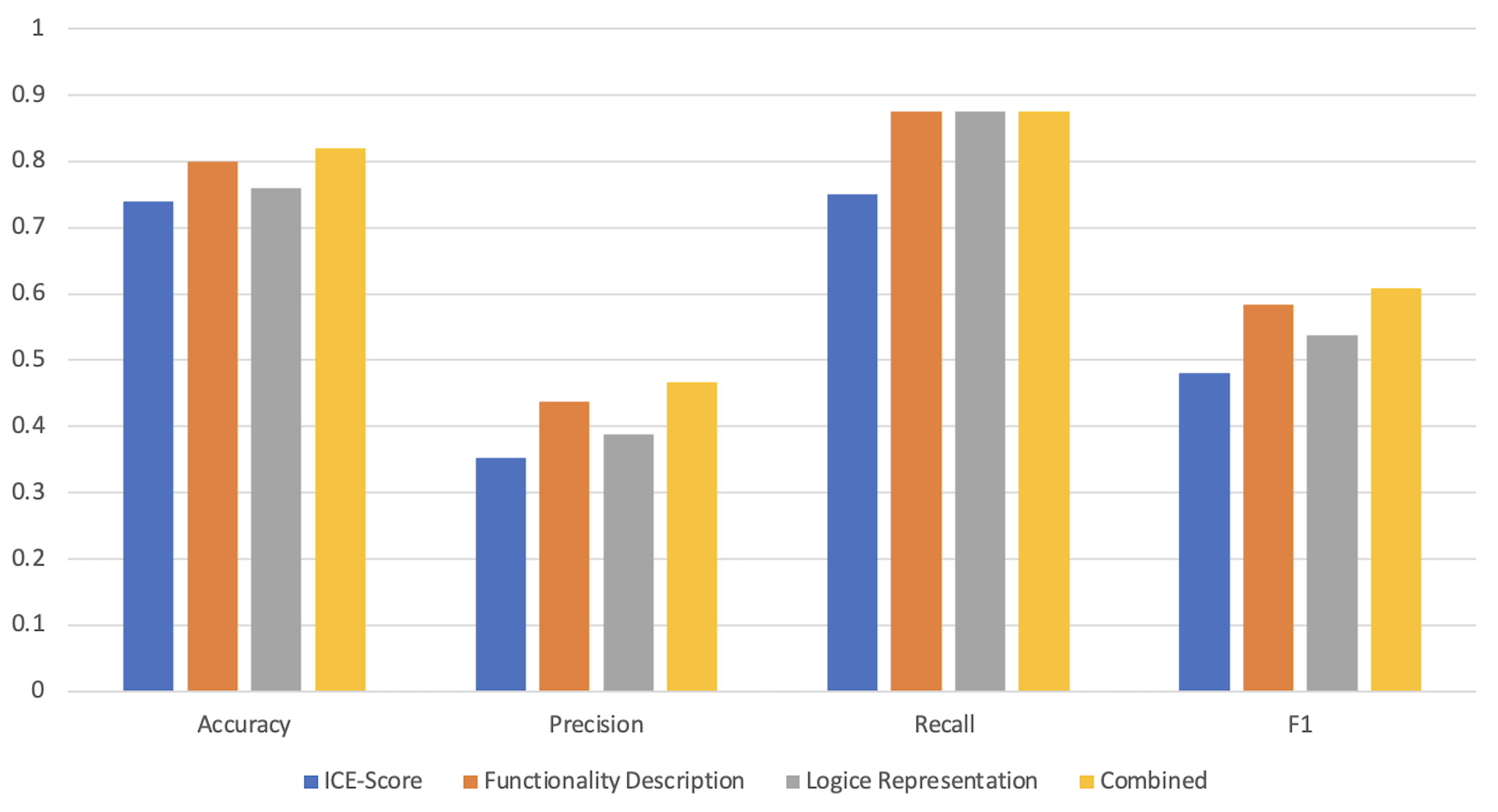}}
\caption{Comparison among LLM-as-a-Judge Metrics for Bash-3}
\label{fig-bashscripts}
\end{figure*}

\section{Evaluations}
\label{sec:app}

\paragraph{Automatic Code Validation} The evaluation of our \texttt{Bidirectional functionality matching} and \texttt{Logic representation} metrics against the baseline \texttt{ICE-Score} using execution-based evaluation as ground-truth (Figures \ref{fig-bash1}, \ref{fig-bash2}, \ref{fig-bashscripts}) shows the following:

\begin{itemize}
    \itemsep0em
    \item Our metrics achieved high agreement on binary judgments with the execution-based evaluation ground-truth.
    \item Typically, our metrics significantly outperformed the baseline on all Accuracy, Precision, Recall, and F1 across three test suites Bash-1, Bash-2, and Bash-3.
    \item The \texttt{Bidirectional functionality matching} metric tends to yield slightly higher performance than \texttt{Logic representation}.
    \item We combined our two metrics as the third one, \texttt{Combined}, and it often achieved slight improvements. 
\end{itemize}

\paragraph{Automatic Code Refinement} To evaluate the automatic code refinement task, we compared our approach (\texttt{Reflection agent with dedicated Evaluator} utilizing judgments and feedback from our two LLM-as-a-Judge metrics) with two different settings: 1) \texttt{non-agentic} as baseline, and 2) \texttt{Self-Reflection agent} using instructions for self-reflection with averaging three iterations. All experiments are zero-shot on 150 questions of three test suites. We used the execution-based evaluation Accuracy metric for comparison.

Our observations from the evaluation results (Table \ref{tab:refine}) are as follows:

\begin{itemize}
    \itemsep0em
    \item Generally, \texttt{Self-Reflection} improved the code quality over baseline.
    \item Most of the time, \texttt{Reflection with dedicated Evaluator} achieved significant improvements over both baseline and \texttt{Self-Reflection}.
    \item Using judgment and feedback from our LLM-as-a-Judge metrics helps to boost the code quality through the evaluation and refinement process but not always; it still depends on the specific model and dataset. Large models tend to follow instructions and improve better (e.g. GPT-4o and Mistral-large vs. Llama-3.3-70B).
    \item Regarding improvement margin using \texttt{Reflection with dedicated Evaluator}, a very large model like GPT-4o may not always obtain the best improvement since it usually performs very well with non-agentic baseline. For instance, Mistral improved the most on Bash-1 over baseline and \texttt{Self-Reflection} (4\% and 4\%, respectively), Llama did best on Bash-2 (24\% and 12\%, respectively), and GPT-4o achieved the highest improvement on Bash-3 (6\% and 2\%, respectively).
    \item Agentic setting is not magic; it often improves performance but occasionally degrades it compared to non-agentic.
    
\end{itemize}

\begin{table}
  \centering
  \begin{tabular}{l|c|c|c}
    \textbf{Models \& Settings} & \textbf{Bash-1} & \textbf{Bash-2} & \textbf{Bash-3} \\
    \hline
    GPT-4o (baseline) & 86\% & 72\% & 68\% \\
    GPT-4o (Self-Ref) & 84\% & 76\% & 72\% \\
    GPT-4o (\textbf{Ref+Eval}) & \textbf{88\%} & \textbf{82\%} & \textbf{74\%} \\
    \hline
    Llama (baseline) & 68\% & 52\% & 46\% \\
    Llama (Self-Ref) & \textbf{74\%} & 64\% & \textbf{52\%} \\
    Llama (\textbf{Ref+Eval}) & 72\% & \textbf{76\%} & 50\% \\
    \hline
    Mistral (baseline) & 76\% & 66\% & 54\% \\
    Mistral (Self-Ref) & 76\% & 70\% & \textbf{58\%} \\
    Mistral (\textbf{Ref+Eval}) & \textbf{80\%} & \textbf{76\%} & \textbf{58\%} \\
  \end{tabular}
  \caption{Execution-based Evaluation Accuracy for Automatic Code Refinement}
  \label{tab:refine}
\end{table}

\section{Conclusion}
Given recent advances of Large Language Models (LLMs), the LLM-as-a-Judge paradigm employing LLMs for automated evaluation has emerged. Given strong coding abilities and reasoning skills, LLMs have the potential to be cost-effective and scalable surrogates for human evaluators. However, this research direction is still in its early stages. In this paper, in the light of automatically selecting the best model and refining code snippets for automatic incident remediation pipeline in IT Automation, we introduced two LLM-as-a-Judge metrics: \texttt{Bidirectional functionality matching} and \texttt{Logic representation}. They proved useful in both tasks of automatic code validation and refinement for NL2Bash. For future work, we will investigate adopting and validating these metrics for other programming languages.

\section*{Limitations}
Since our work relies on several advanced capabilities of LLMs, it has the following limitations:
\begin{itemize}
    \itemsep0em
    \item the ability of LLMs to generate correct functionality descriptions for the given problem and the code snippet.
    \item the reasoning ability of LLMs to align and match the two functionality descriptions of the given problem and the code snippet.
    \item the ability of LLMs to translate the code snippet into a correct logic representation.
    \item the reasoning ability of LLMs to compare if the logic representation of the code snippet satisfies and covers all required functionalities of the given problem.
    \item generally, our work assumes that LLMs can understand and follow instructions to accurately execute steps for automatic code validation and refinement.
    \item as for automatically extracting code snippets from LLMs' outputs, sometimes LLMs do not follow instructions and return long texts consisting of code, explanation, reasoning, examples and hallucination. These require manual extraction of the code snippets from the LLMs' outputs.
\end{itemize}

\bibliography{custom}

\appendix

\section{Appendix}
\label{sec:appendix}

\begin{figure*}[htbp]
\centerline{\includegraphics[scale=0.87]{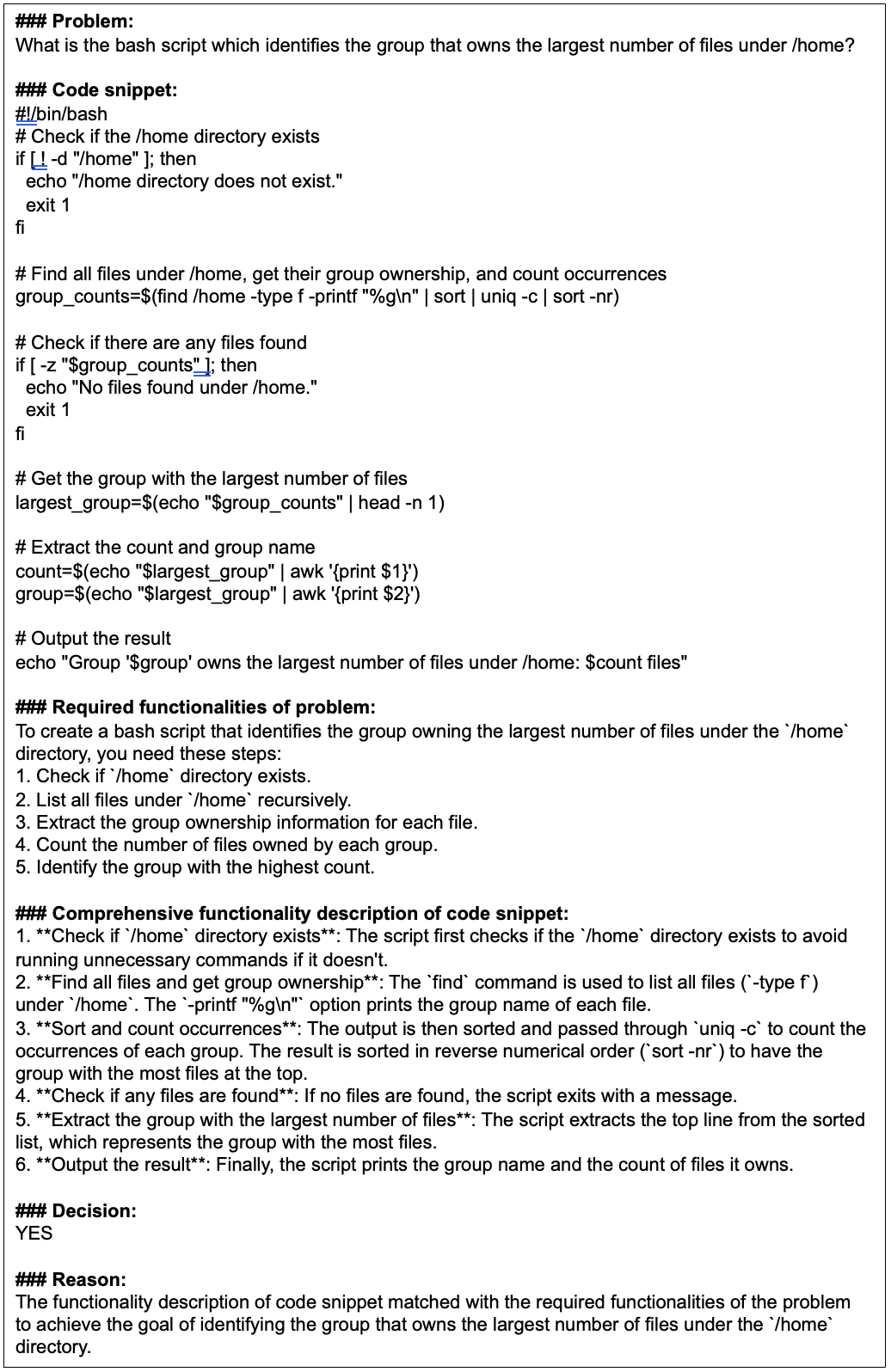}}
\caption{Example of Bidirectional Functionality Matching Validation}
\label{fig-func-des}
\end{figure*}

\begin{figure*}[htbp]
\centerline{\includegraphics[scale=0.85]{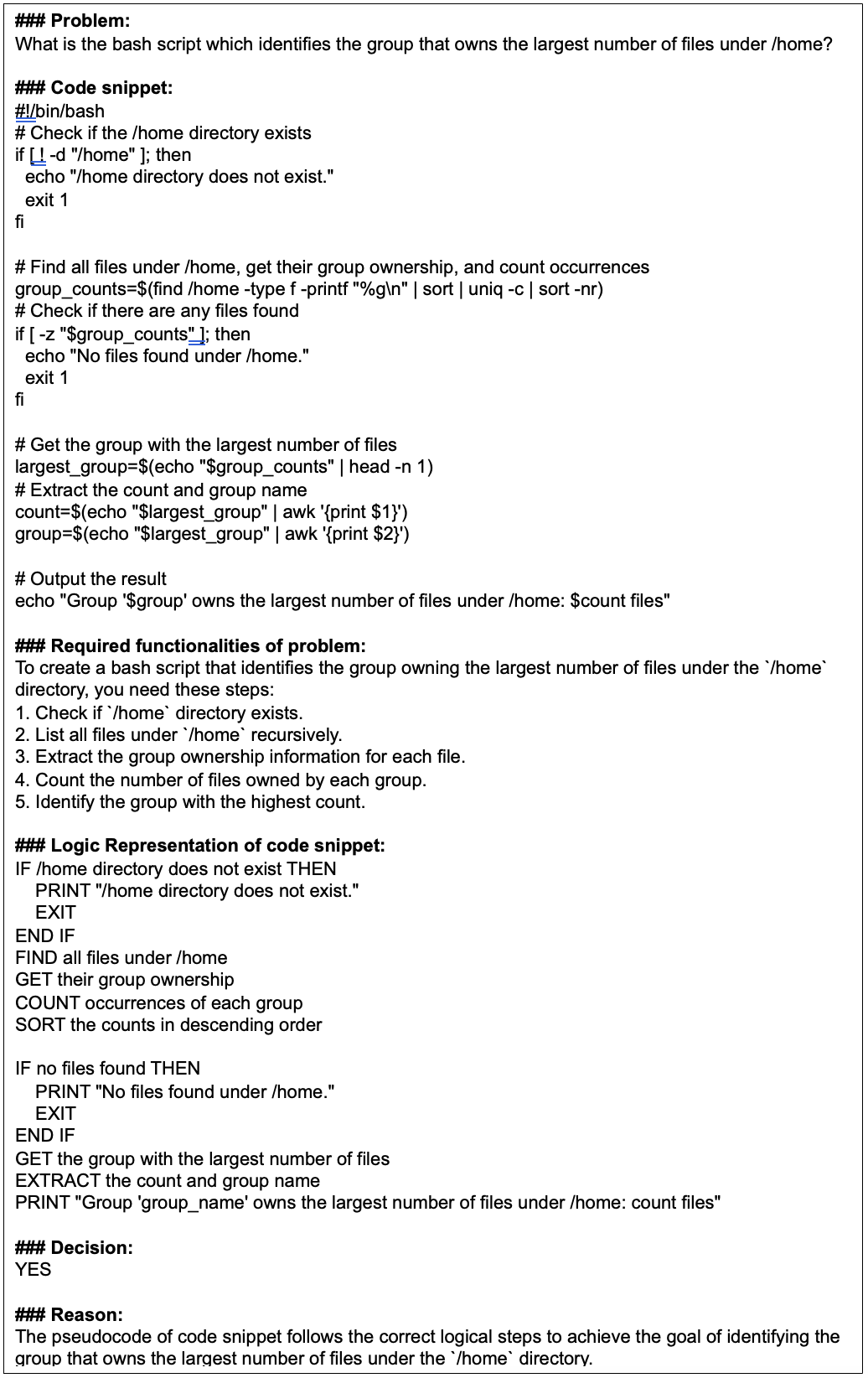}}
\caption{Example of Logic Representation Validation}
\label{fig-logic-rep}
\end{figure*}

We have included two examples of how \texttt{Bidirectional functionality matching} and \texttt{Logic representation} metrics work for automatic code validation (Figures \ref{fig-func-des} and \ref{fig-logic-rep}).

% For example:\\
% Task: What is the bash command which lists the name and pids of the top 5 processes by percentage of CPU usage?
% Input: You are a meticulous Bash programming expert. Please generate a precise and concise functionality description for the following Bash script:
% ```
% \#!/bin/bash
% top -b -n 1 | head -n 11 | tail -n 5 | awk '{print $11, $1}'
% ```

% Output:
% This Bash command lists the names and PIDs of the top 5 processes by CPU usage percentage. Here's a breakdown of the command:
% 1. `top -b -n 1`: Runs the `top` command in batch mode and updates the output once.
% 2. `head -n 11`: Extracts the first 11 lines of the `top` command's output.
% 3. `tail -n 5`: Selects the last 5 lines from the previous output, which correspond to the top 5 processes by CPU usage.
% 4. `awk '{print $11, $1}'`: Prints the 11th field (command name) and 1st field (PID) from the selected lines.
% The final output consists of the command names and PIDs of the top 5 CPU-consuming processes.

\end{document}